\documentclass{vgtc}                          

\graphicspath{{figures/}{pictures/}{images/}{./}} 

 \usepackage{times}                     

  \usepackage{booktabs}                  
 \usepackage{lipsum}                    
 \usepackage{mwe}                       

  \usepackage{mathptmx}                  

\onlineid{0}

\vgtccategory{Research}

\vgtcinsertpkg

\title{Identifying Locally Turbulent Vortices within Instabilities}

\author{Fabien Vivodtzev\thanks{e-mail: fabien.vivodtzev@cea.fr}\\
        \scriptsize CEA, CESTA
\and Florent Nauleau\thanks{e-mail: florent.nauleau@cea.fr}\\
     \scriptsize CEA, CESTA
\and Jean-Philippe Braeunig\thanks{e-mail: jean-philippe.braenig@cea.fr}\\
     \scriptsize CEA, CESTA     
\and Julien Tierny\thanks{e-email: julien.tierny@sorbonne-universite.fr}\\
  \scriptsize CNRS, Sorbonne Université, LIP6.}

\teaser{
  \centering
  \vspace{-3ex}
  \includegraphics[width=\linewidth]{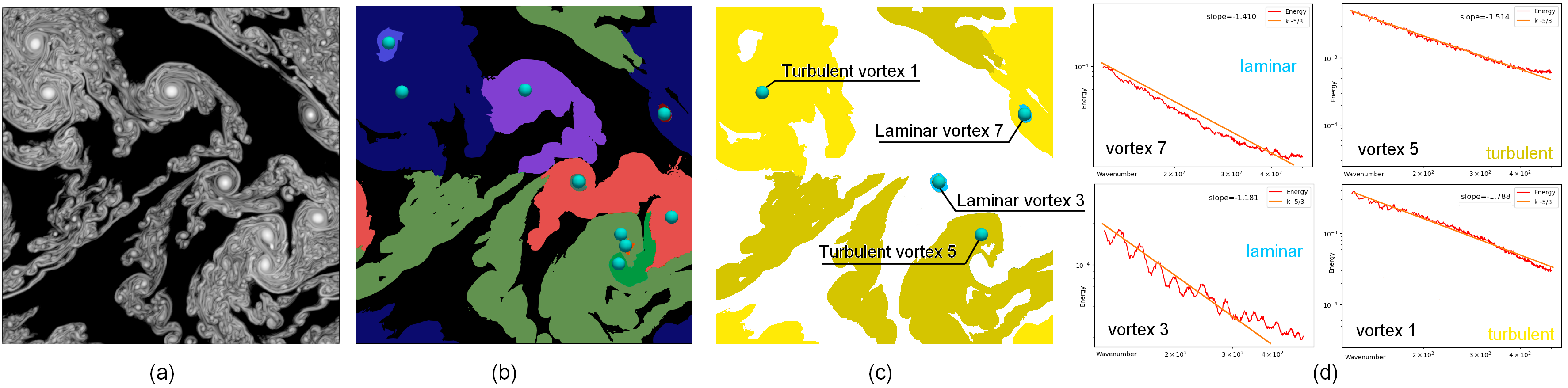}
  \vspace{-3ex}
  \caption{Flow frequency analysis of topologically segmented vortices : $(a)$ 2D view of the Enstrophy $(b)$ Morse-Smale segmentation (after persistence simplification).
  $(c)$ Classification of four segmented vortices as laminar or turbulent made by CFD experts.  $(c)$ Kinetic energy spectrum computed on each classified vortex used to derive an indicator of local turbulence.}
  \label{fig:teaser}
}

\abstract{
    This work presents an approach for the automatic detection of locally 
turbulent vortices within turbulent 2D flows such as instabilites. First, given a time step of 
the flow, methods from Topological Data Analysis (TDA) are leveraged to extract 
the geometry of the vortices. Specifically, the enstrophy of the flow is 
simplified by topological persistence, and the vortices are extracted by 
collecting the basins of the simplified enstrophy's Morse complex. Next, the 
local kinetic energy power spectrum is computed for each vortex. We introduce a 
set of indicators based on the kinetic energy power spectrum to estimate the 
correlation between the vortex's behavior and that of an idealized turbulent 
vortex. Our preliminary experiments show the relevance of these 
indicators for distinguishing vortices which are turbulent
from those which have not yet reached a turbulent state and thus known as laminar.
} 

\begin{document}


\firstsection{Background}

\maketitle

This work focuses on providing a better understanding of vortices in the context of numerical simulation of high-speed flows where turbulence may occur. 
Preliminary studies have been conducted to motivate this work. 
\cite{bridel_ldav19} showed that existing TDA methods could be used to provide an analysis of high velocity compressible turbulent flow. 
Several aspects of the flow have been correlated with TDA structures, such as the location of wavemakers or the frequency of vortex shedding. 
A more complete experimental evaluation of TDA is described in \cite{nauleau_ldav22}. Computed Fluid Dynamic (CFD) hypotheses have been correlated to TDA pipelines (using persistence, Wasserstein distance, clustering ...) to help scientists compare different numerical methods implemented in the simulation codes (interpolation, solvers, orders ...). 
These previous studies motivated this work to better show the relevance of the feature extracted with TDA with respect to traditional turbulence descriptors such as those introduced in Kolmogorov work detailed in section \ref{turbulence}.

\section{Numerical simulation}
A CFD simulation code \cite{bridel2021immersed} was used to generate the data used in this work. It is based on the two-dimensional compressible unsteady Euler equations for inviscid flows. It is a massively parallel structured solver with immersed boundary conditions.
To solve the Euler equations numerically, a Riemann problem is defined at the interfaces between the cells of the mesh. 
The approximate Riemann solver \cite{toro2013riemann} AUSM$^+$-UP \textit{(Advection Upstream Splitting Method +UP)} \cite{liou2006sequel} is implemented to reconstruct strong discontinuities.
To emulate turbulence in an infinite medium, all boundary conditions are set to periodic.
Finally, to capture the turbulence in two dimensions in a scalar, we rely on the local enstrophy $\mathcal{E}$, locally defined as the square of the flow vorticity.

\section{Topological Data Analysis}
Topological data analysis is a set of techniques \cite{edelsbrunner09, tierny_book} that focus on structural features in data. 
For our analysis, we used several well-established techniques that are readily available in the Topology ToolKit (TTK) \cite{ttk17, ttk19, leguillou_tvcg24}. 
The input data are given as an ensemble of piecewise linear scalar fields.
In the case of flow enstrophy, critical points representing local maxima denote the center of vortices in the turbulent flow. 
Because of their locality, a slight oscillation in the data leads in practice to the appearance of spurious critical points, especially in the case of noisy data such as turbulent flows. 
Therefore, the notion of persistence is used to evaluate the importance of a critical point. 
Persistence pairs of critical points can be represented with persistence diagrams. 
The Morse-Smale complex is a topological data structure that provides an abstract representation of the gradient flow behavior of a scalar field. It subdivides a given scalar field
into regions of uniform gradient flow, segmenting the domain such that every point in the same ascending manifold flows toward the same critical point.
In this work, we threshold each ascending manifold above an arbitrarily low threshold of the original  enstrophy (typically $10^{-6}$), such that the boundaries of the extracted regions match the lowest level sets of enstrophy.

\section{Turbulence characterization}
\label{turbulence}
\begin{figure}
  \centering
  \includegraphics[width=\linewidth]{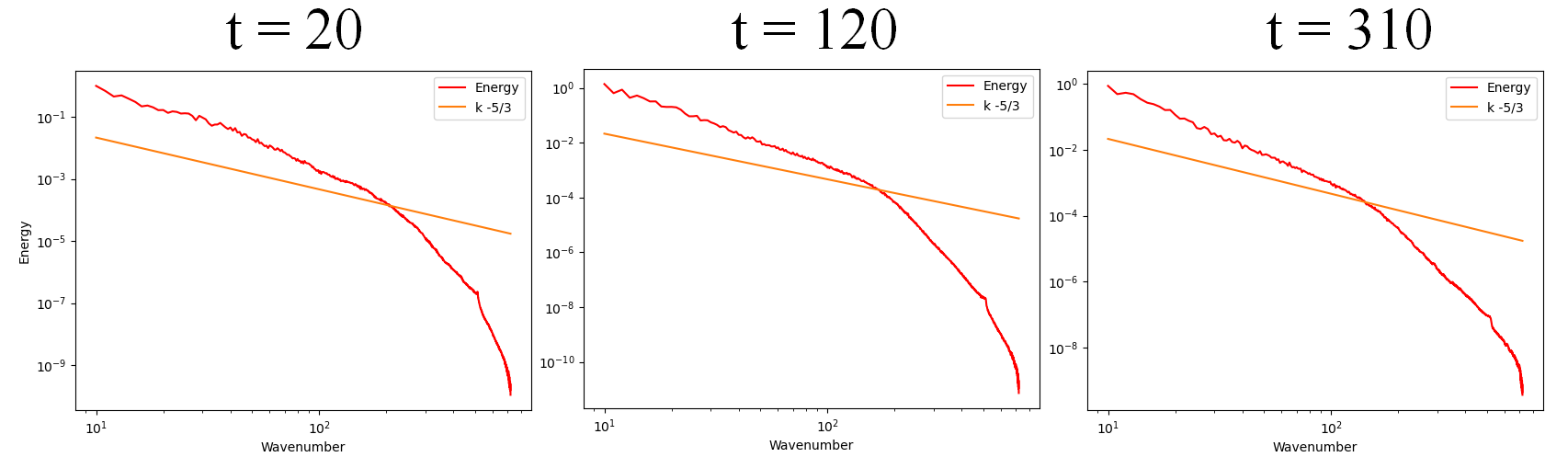}
  \caption{Energy spectrums of turbulence computed on 3 time steps $(t=20$, $t=120$, $t=310)$ showing the energy cascade from low wavenumbers to the highest (orange line: $k^{-5/3}$ linear fit).}
  \label{fig:spectrum}
\end{figure}

In a typical turbulent flow, there is a wide range of vortex sizes that fluctuate at different frequencies.
In a laminar flow, the flow velocity varies smoothly and predictably in space and time, whereas a turbulent flow exhibits a chaotic behavior. 
Turbulence is characterized by energy transfers between these small and large eddies. 
In the 1940s, Andrey Kolmogorov introduced several statistical properties for these flows for an intermediate range of eddy scales, called the inertial subrange. The one used in this work is the energy spectrum of turbulence $E(k)$ related to the mean turbulence kinetic energy as a function of wavenumber ($a.k.a$ the spatial frequency) $k$ as shown in Figure \ref{fig:spectrum}. 
In the intertial subrange, the energy transfer from low to high wavenumber can be described \cite{obukhov41} as the uniform form of E(k) $\sim k^{-5/3}$.   
In this work, the characterization of turbulence in the topological structures extracted by our pipeline is based on this concept. The kinetic energy spectrum is computed via a Fast Fourrier transform and radial averaging \cite{NAVAH2020112885}.  The slope of the linear regression of the energy spectrum, in the intertial subrange, is used as an indicator of turbulence. Other indicators are also tested, such as the minimum and maximum of $E(k)$.



\section{Case Study on a turbulent flow}
The initialization of the turbulence was generated with two fluids of different densities and different velocities of opposite direction, creating a shear zone where the turbulence appears. 
The boundaries are set to be periodic. 
A total of 421 snapshots were taken to capture the evolution of the turbulence on Cartesian grids of $1024 \times 1024$. 
A topological simplification is performed on the enstrophy scalar field based on a threshold of $0.3$ on the (normalized) topological persistence. Then the vortices are extracted by collecting the basins of the ascending Morse-Smale complex of the simplified enstrophy. In this case study, the interval between wavenumbers 120 and 500 is represented, which we use to compute our indicator (slope of the linear regression) in this interval.

\begin{figure}
\centering
  \includegraphics[width=\linewidth]{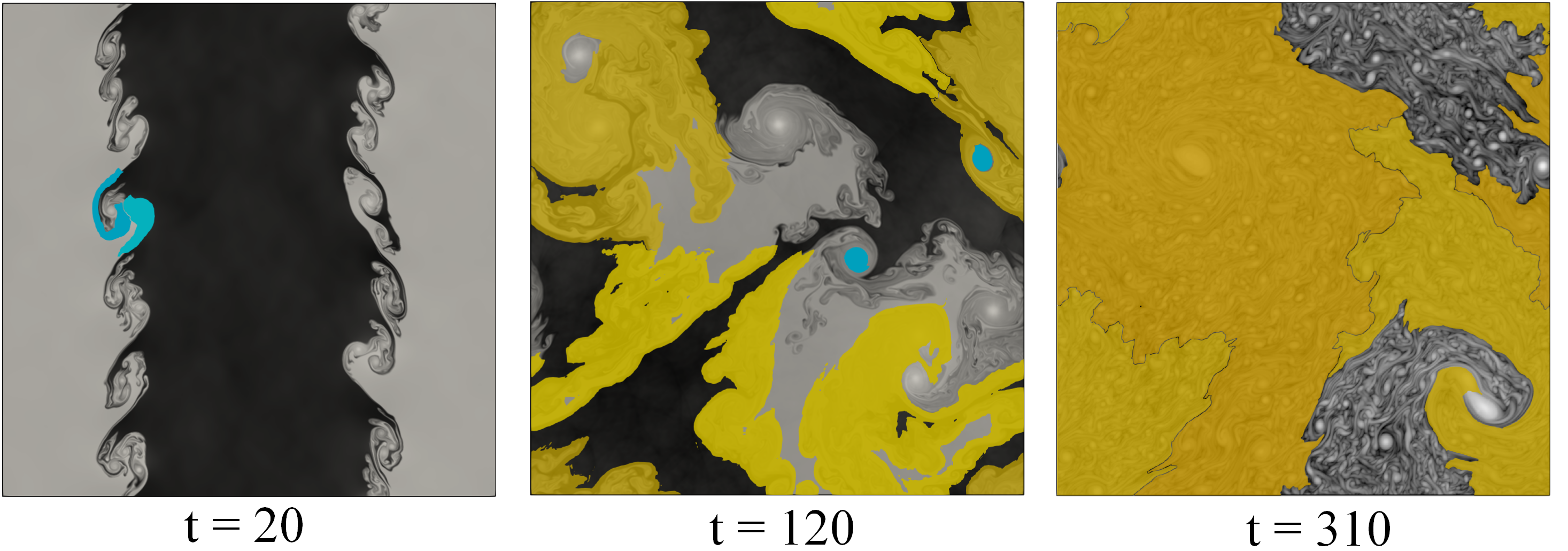}
  \caption{Ground-truth classification with laminar vortices in blue and turbulent vortices in orange annotated by CFD experts.}
  \label{fig:vortices}
\end{figure}

\section{Interpretation}
A classification of the segmented vortices based on the Morse-Smale complex was performed by CFD Expert as shown in Figure \ref{fig:vortices}, which provides a ground truth description of the flow (laminar vortices in blue and turbulent vortices in orange).
The kinetic energy spectrum was calculated for each segmented vortex. Then several indicators are evaluated, such as the slope of the linear regression on the intertial subrange (wavenumbers between 120 and 500 in this case study) or the minimum and maximum of the kinetic energy. Table \ref{fig:indicators} summarizes these indicators for the vortices segmented in Figure \ref{fig:vortices}. It successfully shows that TDA segmented vortices identified as turbulent by CFD experts respect the property E(k) $\sim k^{-5/3}$ described in section \ref{turbulence}.

\begin{table}
\centering
  \caption{Statistical indicators computed on segmented topological features. 
  Vortex geometries
  are topologically segmented and the slope, $E_{max}(k)$ and $E_{min}(k)$ are computed. Vortices with a slope close to $-5/3$ are successfully identified as turbulent, while others are identified as laminar.}
  \includegraphics[width=\linewidth]{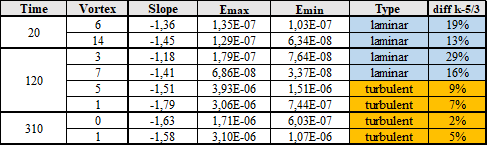}
  \label{fig:indicators}
\end{table}


\section{Conclusion}
In this work, we showed as illustred on Figure \ref{fig:teaser} how topological data analysis could be combined with flow frequency analysis for the detection of vortices which have reached a turbulent state. 
Our preliminary experiments demonstrate the relevance of our new indicators for distinguishing vortices which have reached a turbulent state from those which have not yet. 
In the future, we plan to extend these promising results within a larger study, leveraging supervised classification methods (based on our new descriptors) for the systematic 
detection of turbulent vortices, with applications in vehicle geometry optimization within high velocity flows.

\acknowledgments{
\footnotesize{
This work is partially supported by the European Commission
grant ERC-2019-COG \emph{``TORI''} (ref. 863464,
\url{https://erc-tori.github.io/})}}

\bibliographystyle{abbrv-doi}

\bibliography{template}
\end{document}